\documentclass{emulateapj}
\usepackage{apjfonts}
\usepackage{epsfig}

\newcommand{\etal}{et al.}

\def\gsim{\lower 2pt \hbox{$\, \buildrel {\scriptstyle >}\over
{\scriptstyle \sim}\,$}}
\def\lsim{\lower 2pt \hbox{$\, \buildrel {\scriptstyle <}\over
{\scriptstyle \sim}\,$}}

\def\chandra{{\sl Chandra}}
\def\xmm{{\sl XMM-Newton}}

\def\cvi{C~{\scriptsize VI}}

\def\ciii{C~{\scriptsize III}}
\def\civ{C~{\scriptsize IV}}

\def\nvii{N~{\scriptsize VII}}
\def\nv{N~{\scriptsize V}}

\def\oviii{O~{\scriptsize VIII}}
\def\ovii{O~{\scriptsize VII}}
\def\ovi{O~{\scriptsize VI}}
\def\oi{O~{\scriptsize I}}
\def\oii{O~{\scriptsize II}}

\def\neviii{Ne~{\scriptsize VIII}}
\def\neix{Ne~{\scriptsize IX}}

\def\hi{H~{\scriptsize I}}

\def\siiv{Si~{\scriptsize IV}}
\def\siiii{Si~{\scriptsize III}}

\RequirePackage[T1]{fontenc} 

\shortauthors{Yao \etal}
\shorttitle{Detecting the WHIM through X-ray absorption lines}

\slugcomment{\em Accepted for publication in ApJ}
\begin{document}

\title{Detecting the warm-hot intergalactic medium through X-ray absorption 
  lines}
\author{Yangsen Yao\altaffilmark{1}, 
	J. Michael Shull\altaffilmark{1}, 
	Q. Daniel Wang\altaffilmark{2},
and
	Webster Cash\altaffilmark{1}} 
\altaffiltext{1}{Center for Astrophysics and Space Astronomy,
Department of Astrophysical and Planetary Sciences,
University of Colorado, 389 UCB, Boulder, CO 80309; yaoys@colorado.edu}
\altaffiltext{2}{Department of Astronomy, University of Massachusetts,
  Amherst, MA 01003}


\begin{abstract}

The warm-hot intergalactic medium (WHIM) at temperatures $10^5-10^7$ K is
believed to contain 30-50\% of the baryons in the local
universe. However, all current X-ray detections of the WHIM at redshifts
$z>0$ are of low statistical significance ($\lsim3\sigma$) and/or
controversial.  In this work, we aim to establish the detection limits of
current X-ray observatories and explore requirements for next-generation
X-ray telescopes for studying the 
WHIM through X-ray absorption lines. We analyze all available grating
observations of Mrk~421 and obtain spectra with signal-to-noise ratio 
(S/N) of $\sim90$ and 190 per 50 m\AA\ spectral bin from 
\chandra\ and \xmm\ observations, respectively. Although these spectra are
two of the best ever collected with \chandra\ and \xmm, 
we cannot confirm the two WHIM systems 
reported by Nicastro et al. in 2005. Our bootstrap simulations
indicate that spectra with such high S/N can\emph{not} constrain the
WHIM with \ovii\ column densities $N_{\rm OVII}\approx10^{15}~{\rm cm^{-2}}$ 
(corresponding to an equivalent widths of 2.5 m\AA\ for a Doppler velocity
of $50~{\rm km~s^{-1}}$) 
at $\gsim3\sigma$ significance level. The
simulation results also suggest that it would take $>60$ Ms for \chandra\
and 140 Ms for \xmm\ to measure the 
$N_{\rm OVII}$ at $\ge4\sigma$ from a spectrum of a background QSO with
flux of $\sim0.2$ mCrab
(1 Crab = $2\times10^{-8}~{\rm erg~s^{-1}~cm^{-2}}$ at 0.5-2 keV). 
Future X-ray spectrographs need
to be equipped with spectral resolution $R\sim4000$ and effective
area $A\ge100~{\rm cm^2}$ to accomplish the similar constraints with an
exposure time of $\sim2$ Ms and would require $\sim11$ Ms to
survey the 15 QSOs with flux $\gsim0.2$ mCrab along which clear intergalactic
\ovi\ absorbers have been detected.

\end{abstract}



\section{Introduction }
\label{sec:intro}

Identifying the ``missing baryons'' is one of the major tasks of modern
cosmology. Observations of the 
microwave background (e.g., \citealt{kom11}) and 
the big bang nucleosythesis model combined with the measurement of 
deuterium abundance (e.g., \citealt{bur01, ome06, pet08}) 
are converging on the cosmological baryon density of 
$\Omega_b=0.0455\pm0.0028$. 
At high redshift ($z>3$) universe, baryons
exist primarily in the form of cool, photoionized intergalactic medium
(IGM) that is traced
by Ly$\alpha$ absorption forest lines \citep{rau97}. In the present-day
universe, matter detected in forms of photoionized IGM, stars, 
galaxies, intracluster medium, etc., adds up to only $\sim50\%$ of the 
baryons \citep{shu11}, 
leaving much of the inventory yet to be found
\citep{fuk98, fuk04}. Cosmological numerical simulations for large-structure
formation indicate that the missing baryons are still in the IGM, but at
current epoch they have been heated by gravitational shocks and galactic
feedback to temperatures of $10^5-10^7$ K when they fall into the
gravitational potential wells of the dark matter cosmic web filaments
(e.g., \citealt{cen99, cen06, dav01}). 
The IGM at these temperatures, so-called warm-hot intergalactic medium
(WHIM), mainly absorbs and emits photons in the 
ultraviolet (UV) and X-ray wavelength bands.

Searching for the missing baryons has been conducted extensively in the 
UV band. Indeed, besides the cool phase Ly$\alpha$ absorbers 
(e.g., \citealt{wey98, pen00, pen04}), the absorption lines of \ciii, \civ,
\nv, \ovi, \siiii, and \siiv\ 
have been routinely observed in spectra of background quasi-stellar objects
(QSOs; e.g., \citealt{tri00, tri08, pro04, leh06, dan05, dan08, dan06,
  thom08, yao11}),  
suggesting that a significant
amount of baryons exist in the highly ionized absorbers. 
These high ions, \ovi\ in particular,  are believed to trace the
WHIM at the low-temperature end ($\sim10^5-10^6$ K). However, all these 
ions can also  
be produced through photoionization in the intergalactic 
environment (\citealt{opp09}; but see also \citealt{tep11} and
\citealt{smi12} for a contrary view)
and thus some of them may not contribute to the canonical WHIM
(i.e., the shock-heated IGM expected from simulations). How many
of these absorbers originate in the real WHIM is still under debate
\citep{dan08, tri08}. \neviii\ and broad \hi\ Ly$\alpha$
absorbers (BLAs) are complementary and valuable tracers of hot 
absorbing WHIM in UV bands.
However, there are only a few detections with marginal significance of
\neviii\ \citep{sav05, nar09, nar11} and detections of BLAs require
very high signal-to-noise-ratio (S/N) of spectra to resolve broad and
weak signals from the continua \citep{ric06, leh07, sav11, dan10, dan11}.
Converting measurement of these UV absorbers to census of baryonic matter
also depends on ionization fractions and metallicity of the IGM, which in
many absorbers are still poorly known. Nevertheless, BLAs and
\ovi-absorbers, although partially overlapped with photoionized Ly$\alpha$
absorbers, are estimated to contribute an additional $\sim25\%$ to the
baryon inventory at the current epoch (e.g., \citealt{dan08, dan11, shu11}).

X-ray observations of the intergalactic \ovii\ and \oviii\
emission/absorption features could provide essential information 
for establishing the existence of the WHIM and completing the baryon
inventory in the local universe. 
At temperatures $T\gsim10^6$ K, hydrogen and helium are nearly completely 
ionized \citep{gnat07}. 
Without metals, the thermal gas emits/absorbs photons through 
bremsstrahlung, which unfortunately is optically thin even at 
intergalactic scales \citep{bre07}. Because the K-shell transitions of all
elements heavier than lithium and the L-shell transitions of elements
heavier than iron  
lie in X-ray bands \citep{pae03}, metals can greatly increase the emissivity
of the gas. However, because of the density-squared dependence, the WHIM
emission is expected to be weak and difficult to detect. In fact, the IGM
emission has been directly observed only from the dense regions like 
intracluster and intragroup media 
(e.g., \citealt{wang97, mul00, all02, sun09}). 
The WHIM emission signals were also claimed in the angular correlation of the
diffuse X-ray background \citep{sol99, urs06}, but disentangling the real
WHIM signal from the unresolved point sources and local Galactic hot
diffuse gas has been a major uncertainty (e.g., \citealt{gal07}). 

Unlike emission lines, absorption lines
measure the column densities and thus directly sample the total mass 
of the intervening gas along a line of sight. Oxygen is the most abundant
metal element. In contrast of \ovi\ as a minority ionization state,
\ovii\ is the most abundant ion for a gas at  
temperatures of $T\sim10^{5.5}-10^{6.3}$ K and \oviii\ takes
over when $T\gsim10^{6.3}$ K \citep{gnat07}. Thus they have
advantages over \ovi\ in estimating the total baryon contained in highly
ionized absorbers. \ovii\ and \oviii, whose production ionization potentials
are 138.1 eV and 739.3 eV compared to 113.9 eV of \ovi, are also
hard to produce through photoionization in the IGM
environment (e.g., \citealt{cen06a, chen03}).
Theoretical calculations indicate that the \ovii\ column density
is about 10 times higher than \ovi\ in a shock-heated gas 
(i.e., $N_{\rm OVII}\gsim10N_{\rm OVI}$) 
whereas $N_{\rm OVII}\lsim3N_{\rm OVI}$ in a photoionized gas
\citep{fur05}. 
Therefore observations of the IGM \ovii\ and \oviii\ absorption lines are
crucial, not only for constraining the properties of the WHIM at high 
temperature end but also for 
probing the nature of the commonly observed \ovi-absorbers.

Unfortunately, most attempts at searching the X-ray WHIM
absorption lines have been frustrated. \ovii\ and \oviii\ absorption lines 
at $z=0$ have been unambiguously detected in many background QSOs. However, 
all these X-ray absorptions are \emph{inconsistent} with an intergalactic
origin \citep{fang06, bre07b, yao10}, but rather can be attributed to
the Galactic diffuse hot gas 
(e.g., \citealt{yao05, yao07, yao08, yao09, wang05}).
Perhaps the most compelling intergalactic 
result is the detection of \ovii\ absorption
at the redshift of the large galaxy structure Sculptor Wall 
($z_{\rm abs}=0.03$) along the QSO sight line H~2356-309 
($z_{\rm QSO}=0.165$), 
albeit of low significance ($\sim3\sigma$; \citealt{buo09, fang10}). 
However, because of the high number density of galaxies along the sight
line, the absorption may sample the halo gas of one or more intervening 
galaxies with small impact distances ($\lsim50$ kpc; \citealt{wil10}), 
i.e., mimicking the \ovii\ and \oviii\ absorption lines at $z=0$. Therefore,
it may not be representative of the typical WHIM. Furthermore, 
attempts at searching for the similar absorption features at the redshift of 
another large structure, Pisces-Cetus along the same sight line, 
failed \citep{zap10}. All other claimed WHIM \ovii\ and \oviii\ absorptions
have been highly debated. For instance, the detected \oviii\ absorption line
at $z=0.0554$ in \chandra\ spectrum of PKS~2155-304 
\citep{fang02a, fang07} cannot be confirmed by the \xmm\
observations \citep{cag04}, although at nearby redshift a small galaxy
group and the corresponding UV \ovi\ and Ly$\alpha$ absorption lines have
been identified 
\citep{shu98, shu03}. \citet{nic05} obtained a spectrum with unprecedented
S/N during the burst states of Mrk~421 and 
reported two WHIM detections at $z=0.011$ and $z=0.027$. Again,
observations with \xmm\ cannot confirm the  
detections and reported significances have also been questioned
\citep{kaa06, ras07}. \citet{wil10} recently found a galaxy
filament at $z=0.027$, which makes the reported \ovii\ WHIM
detections along the Mrk~421 sight line back to be a subject of discussions.

In this work, we aim to provide detection limits of current X-ray
observatories, \chandra\ and \xmm, in measuring the
WHIM through X-ray absorption lines, and 
establish requirements (e.g., effective area and spectral resolution)
for next-generation X-ray telescopes. We begin our
investigation by scrutinizing the controversial WHIM detections
along the Mrk~421 sight line. We extensively explore all the available
\chandra\ and \xmm\ observations, and then run bootstrap simulations to
test the reliability of any absorption feature.

The paper is organized as follows. In Section~\ref{sec:obs} we describes
the observations and our data reduction processes, and in
Section~\ref{sec:res} we report the data analysis results and compare them
with previous work. In Section~\ref{sec:sim}, we run bootstrap
simulations to explore the detection limits of \chandra\ and \xmm\ and to
establish requirements of next-generation X-ray telescopes. 
We summarize our results in Section~\ref{sec:dis}.

Throughout the paper, we adopt the atomic data from \citet{ver96} and solar
abundances from \citet{asp09}. We conduct our
data analysis within {\sl XSPEC} (version 12.6; \citealt{arn96}), and report
the 1$\sigma$ confidence range or 3$\sigma$ limit 
for a fitting parameter. 
We take the flux normalization of
$1~{\rm Crab}=2\times10^{-8}~{\rm erg~ s^{-1}~cm^{-2}}$ at
0.5-2.0 keV. Thus 0.2 mCrab is 
$2.15\times10^{-13}~{\rm erg~s^{-1}~cm^{-2}}$~\AA$^{-1}$ 
for a flat spectrum in wavelength space, which corresponds to  
$2.34\times10^{-4}~{\rm photons~s^{-1}~cm^{-2}}$~\AA$^{-1}$ around the
\ovii\ K$\alpha$ (21.602 \AA) at the rest-frame. We define measurement 
significance level (SL) of an absorption line as
\begin{equation}
\label{equ:sig}
SL=\frac{EW}{\Delta EW},
\end{equation}
where $EW$ and $\Delta EW$ are the equivalent width and its $1\sigma$ 
uncertainty for an absorption line.


\section{Observations and Data Reduction}
\label{sec:obs}

\begin{figure*}
  \plotone{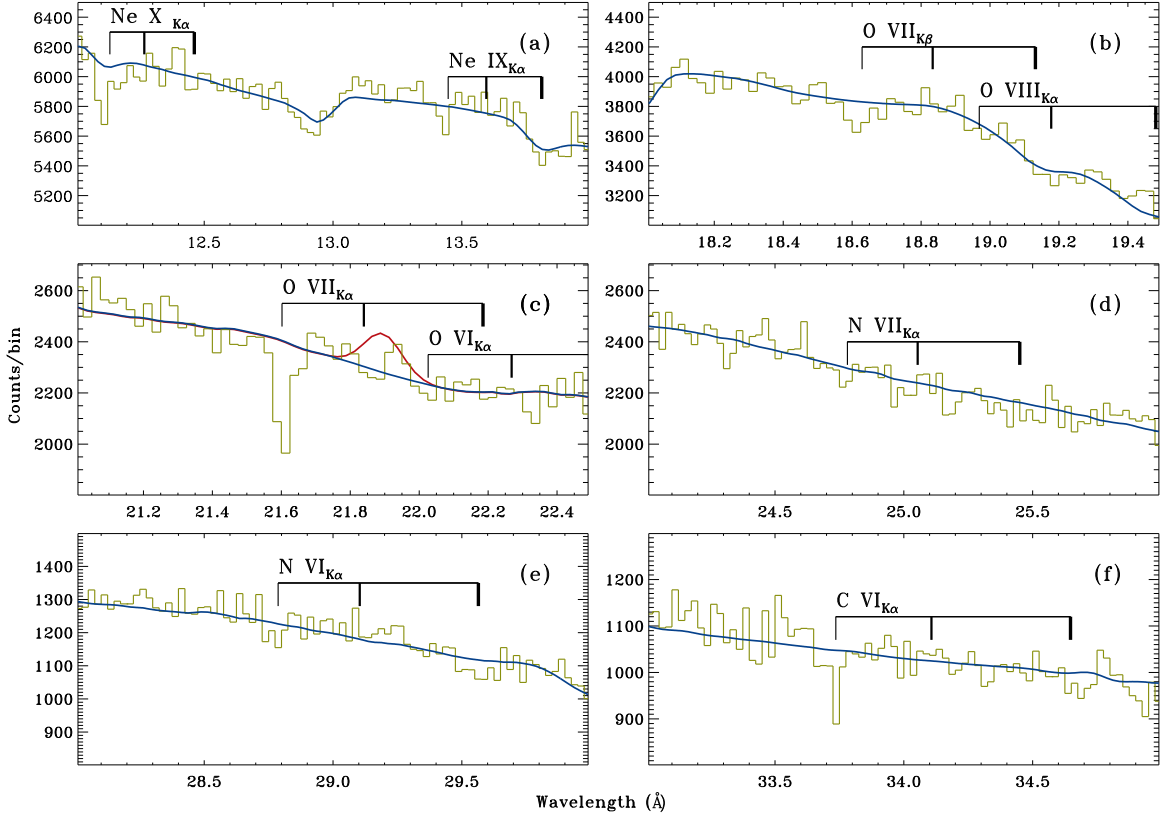}
  \caption{Six portions of coadded {\sl Chandra}-LETG spectrum (green
  histogram) of Mrk~421 with the best-fit continuum model (blue
  curves). The model is convolved with the instrumental line spread
  functions. 
  The binsize is 25 m\AA. The spectrum (Spectrum I) was
  extracted from the same observations as used by \citet{nic05}, and the
  plots reproduce their Figure 8. 
  K$\alpha$ and K$\beta$ transitions of eight ions covered by
  these spectral ranges are labeled in the plots, and the three vertical
  bars from left to right mark the corresponding rest-frame wavelengths
  (thin line), the expected wavelengths at $z=0.011$ (medium line), and
  $z=0.027$ (thick line). The red curve in panel (c) indicates the
  additional component added to local continuum in order to ``amplify''
  the significance of the putative \ovii\ K$\alpha$ absorption line at
  $z=0.011$. The additional component is statistically unnecessary, and the
  red curve is for comparion purpose only. See text in 
  Section~\ref{sec:res} for details.
  } 
  \label{fig:chan-Ni}
\end{figure*}

\begin{figure*}
  \plotone{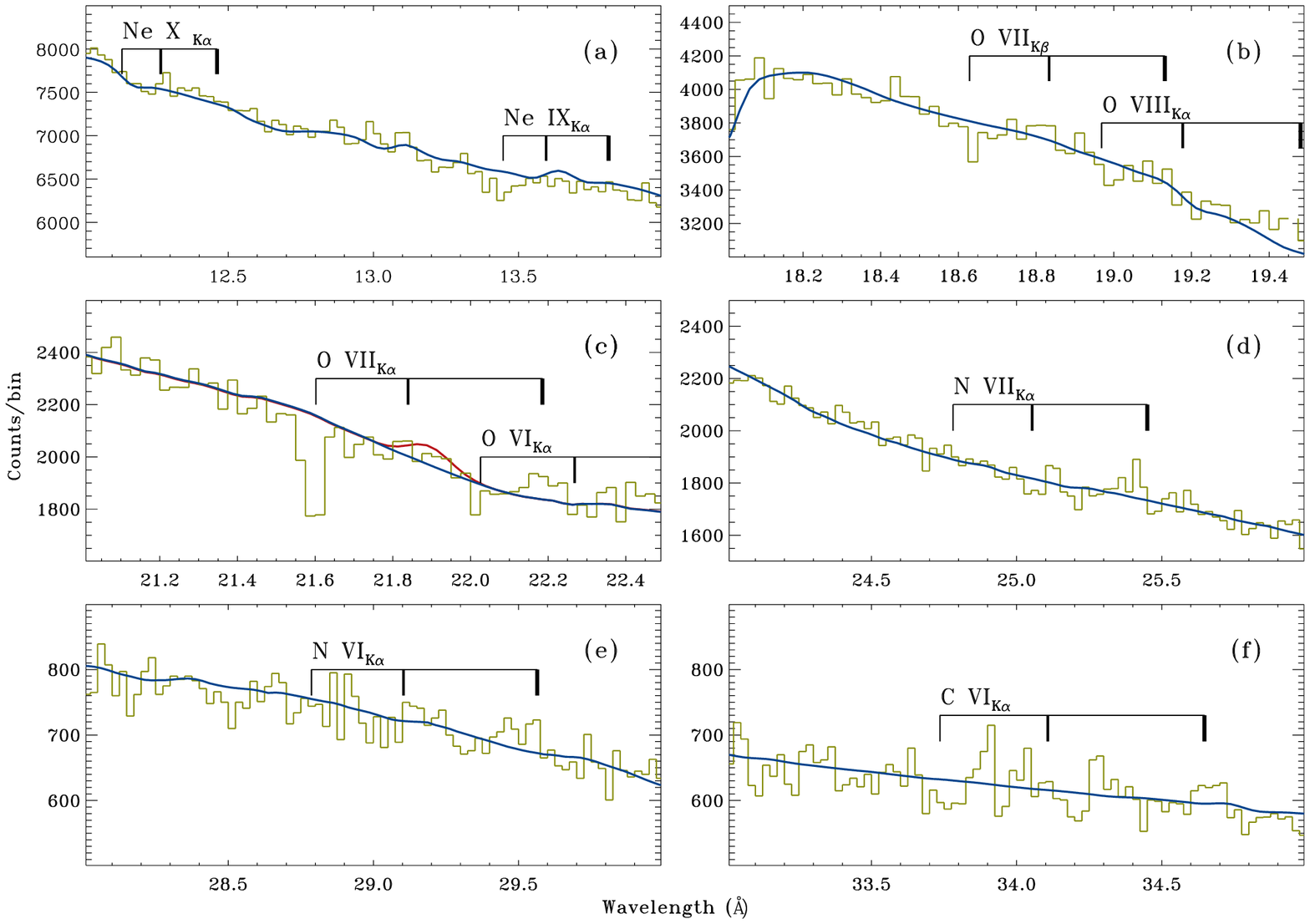}
  \caption{Same as Figure~\ref{fig:chan-Ni}, except that the spectrum 
    (Spectrum II)
  was extracted from the newly available {\sl Chandra}-LETG observations.
  } 
  \label{fig:chan-new}
\end{figure*}

\begin{figure*}
  \plotone{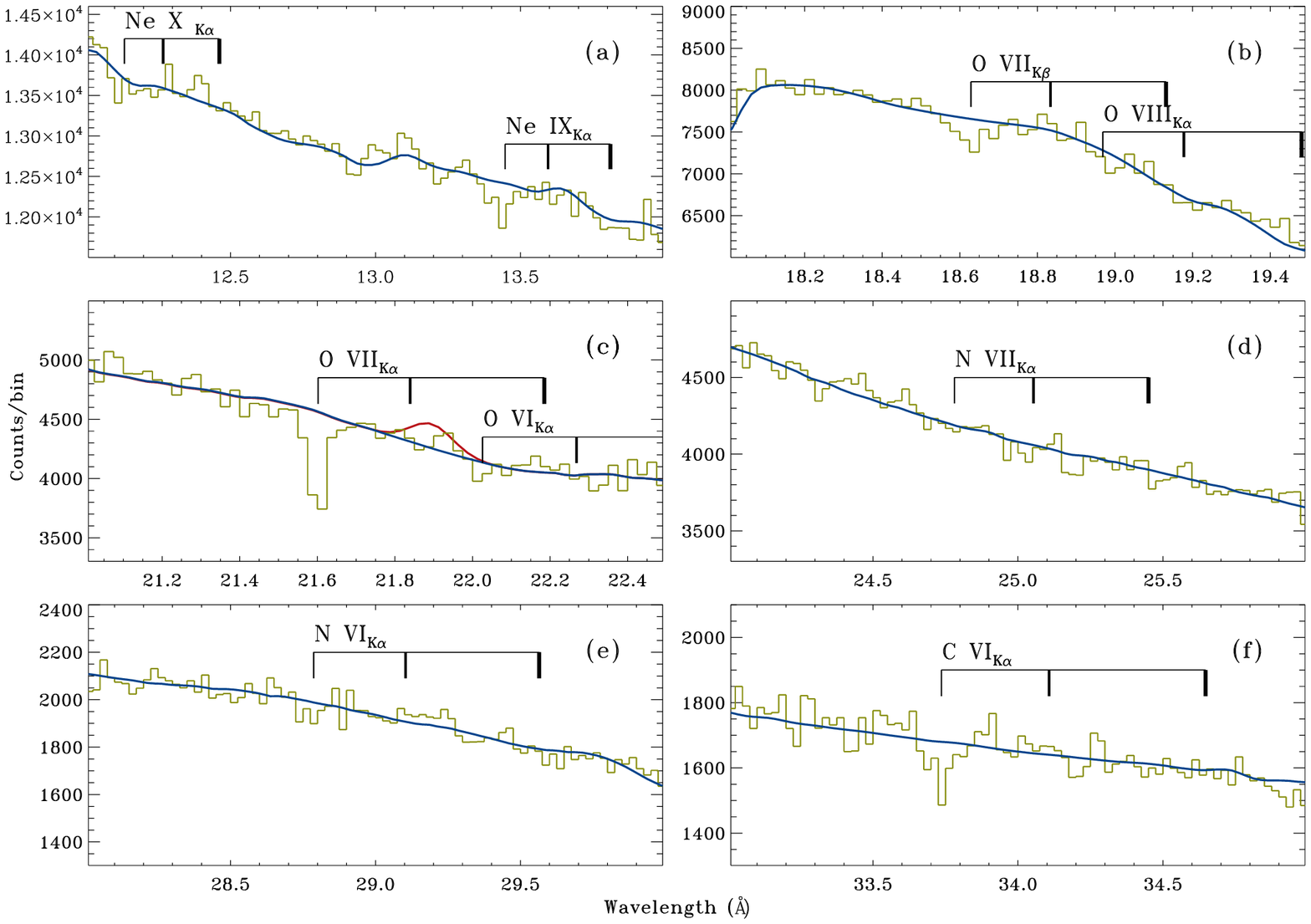}
  \caption{Same as Figure~\ref{fig:chan-Ni}, except that the spectrum
    (Spectrum III)
  was extracted from all available {\sl Chandra}-LETG observations.
  } 
  \label{fig:chan-all}
\end{figure*}

We used all grating observations of Mrk~421 observed with both \chandra\ 
and \xmm\ that are available as of 2011 April. Since Mrk~421 is a
calibration source for both observatories, observations were taken on a 
nearly regular basis. The \chandra\ data
were downloaded from the CXO archive web site
\footnote{http://cda.harvard.edu/chaser/mainEntry.do}, which included 
26 observations taken with the low energy transmission grating (LETG; ObsIDs
of 1715, 4148, 4149, 5171, 5318, 5331, 5332, 6925, 8378, 8396, 
10664, 10665, 10667, 10668, 10669, 10671, 11605, 11606, 11607, 11966, 
11970, 11972, 11974, 12121, 12122, and 13097) with total exposure of 
752.8 ks. This dataset does not include observations taken
with imaging mode and those with the high energy transmission grating (HETG). 
A quick estimate indicates that these excluded observations would
contribute additional $<10\%$ to the total spectral counts, and thus our
results will not be noticeably affected with or without them. For each
employed observation, we used software package CIAO (version 4.3)
\footnote{http://cxc.harvard.edu/ciao/index.html} and followed 
the standard procedure to re-calibrate, extract spectrum, and calculate the
instrumental redistribution matrix file (RMF) and ancillary response file
(ARF). For observations that used the Advanced CCD Imaging
Spectrometer (ACIS) as the detector, we only utilized the first grating
orders. The High Resolution Camera (HRC) does not have spectral
resolution itself, and the spectral orders in observations with HRC
cannot be sorted out. Therefore, for these observations, we calculated
the RMFs and ARFs up to six grating orders and combined them as one in
spectral analysis to account for contributions from
different orders (see descriptions in Section~\ref{sec:res}). To improve
spectral S/N, we utilized the user-developed 
IDL tools \citep{yao07} to combine the negative and positive spectral
orders, coadd spectra of different observations, and calculate the averaged
RMF and ARF by using the spectral counts around 20 \AA\ as weights. The
claimed two WHIM detections were based on a subset 
of the observations listed above (\citealt{nic05};
e.g., ObsIDs of 1715, 4148, and 4149\footnote{\citet{nic05} in fact also
  included two additional HETG observations, which contributed
  negligible spectral counts to their final spectrum.}). 
To scrutinize the reported WHIM
detections, we produced three coadded \chandra\ spectra 
and the corresponding RMFs and ARFs, based upon observations used by 
Nicastro et al. (Spectrum I), observations not included in 
their work (Spectrum II), and 
all observations together (Spectrum III),
respectively. Figures~\ref{fig:chan-Ni}-\ref{fig:chan-all} show six
portions of these spectra.

\begin{figure*}
  \plotone{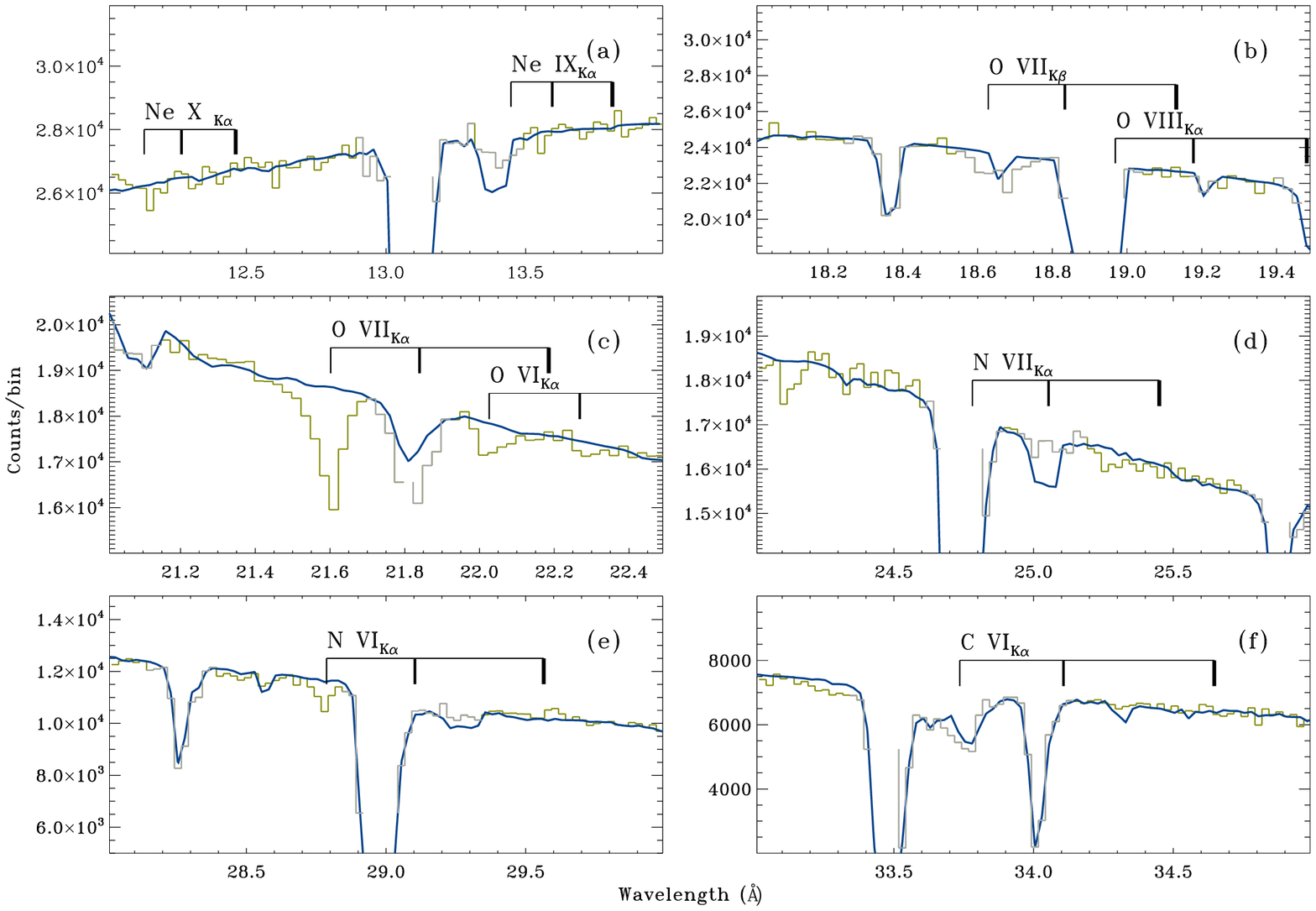}
  \caption{Same as Figure~\ref{fig:chan-Ni}, except that the spectrum
    (Spectrum IV)
    was extracted from all available {\sl XMM-Newton}-RGS observations. 
  Channels/bins colored as grey were contaminated by hot pixels and were
  ignored in our spectral analysis. Panels (b)-(f) were extracted from the
  RGS1 observations, and panel (a) was extracted from the RGS2 observations. 
  } 
  \label{fig:rgs}
\end{figure*}

The total 76 \xmm\ observations were downloaded from the \xmm\ science
archive \footnote{http://xmm.esac.esa.int/xsa/index.shtml.}. We used
script package SAS (version 10.0) \footnote{http://xmm.esac.esa.int/sas/.}
to reprocess these observations. To avoid
the severe contamination from the background flares, 
for each observation we calculated the count rate on the chip CCD9, and
created the good time intervals (GTIs) by excluding the time periods
with rates higher than 0.2 counts per second. After this filter, there are 
54 remaining observations, and each of them has the integrated GTIs longer
than 5 ks. 
We then extracted spectra, calculated RMFs for each of
exposures taken with the reflection grating spectrometer (RGS), 
and combined the spectra of these 54 observations and the corresponding
RMFs by running the script {\sl rgscombine} to get the final coadded
spectrum (Spectrum IV; Figure~\ref{fig:rgs}), 
which has a total exposure of 1.2 Ms.

We used a binsize of 25~m\AA\ to re-bin all the coadded spectra, which is
equivalent to an oversample factor of 2 for a $\sim$50~m\AA\
resolution element of the LETG and RGS. Table~\ref{tab:results} summarizes
the S/Ns of these spectra.

\begin{deluxetable*}{l|l|cr|cr|cr|cr}
\tablewidth{0pt}
\tablecaption{Spectral analysis results \label{tab:results}}
\tablehead{
& & \multicolumn{6}{c}{\chandra} & \multicolumn{2}{c}{\xmm} \\
line & redshift & S/N$^1$ & EW (m\AA)$^1$ & 
                  S/N$^2$ & EW (m\AA)$^2$ & 
                  S/N$^3$ & EW (m\AA)$^3$ & 
                  S/N & EW (m\AA)}
\startdata
\ovii$^4$ & $z=0$     & 68.9 &$10.9\pm0.9$& 64.9& $12.5\pm0.9$ &94.5 & $11.7\pm0.6$ & 193.0 & $14.9\pm0.6$  \\
      & $z=0.011$ & 67.6 &$<3.0$      & 62.8& $<1.5$       &92.2 & $<1.6$       & $\cdots$ & $\cdots$ \\
      & $z=0.027^6$ & 66.2 &$<3.1$      & 60.6& $<1.5$       &89.6 & $<2.0$       & 187.4 & $<1.6$   \\
\hline 
\ovii$^5$ & $z=0$     & 68.9 &$11.3\pm1.0$& 64.9& $11.8\pm1.1$ &94.5 & $11.6\pm0.9$ & $\cdots$ & $\cdots$ \\ 
      & $z=0.011$ & 67.6 &$3.3\pm1.2$ & 62.8& $<2.0$       &92.2 & $2.0\pm1.1$  & $\cdots$ & $\cdots$ \\
      & $z=0.027^6$ & 66.2 &$<3.3$      & 60.6& $<1.5$       &89.6 & $<2.0$       & $\cdots$ & $\cdots$ \\ 
\enddata
\tablecomments{The S/N is given per 50~m\AA\ resolution
  element around the lines, and EW is the measured equivalent width. 
  Error ranges are quoted at $1\sigma$ significance level, while the upper
  limits are at $3\sigma$. \\
  $^1$ Measurements are obtained from the spectrum corresponding to 
  observations used by \citet{nic05} (Spectrum~I; Figure~\ref{fig:chan-Ni}).\\ 
  $^2$ Measurements are obtained from the spectrum corresponding to those
  observations not used by Nicastro et al. (Spectrum~II; 
  Figure~\ref{fig:chan-new}). \\
  $^3$ Measurements are obtained from the spectrum corresponding to all
  observations (Spectrum~III; Figure~\ref{fig:chan-all}). \\
  $^4$ EW measurements in these three rows are obtained from the best-fit 
       continua without compensation of the 11th broad Gaussian 
	(blue curves in Figures~\ref{fig:chan-Ni}-\ref{fig:rgs}).\\
  $^5$ EW measurements in these three rows are obtained from the best-fit
  	continua with compensation of the 11th broad Gaussian
	(red curves in Figures~\ref{fig:chan-Ni}-\ref{fig:chan-all}).
	However, since the 11th broad Gaussian profile is statistically 
 	unnecessary, these measurements are for comparison purpose only. \\
  $^6$ The measured upper limits to EW of the \ovii\ K$\alpha$ are
       obtained by fixing the line centroid at $z=0.028$ instead of
       $z=0.027$, since \citet{nic05} reported \ovii\ line at the 
 	former redshift. \\    
See text for details. }
\end{deluxetable*}

\section{Spectral analyses, results, and discussion}
\label{sec:res}

Our goal in this section is to examine the WHIM detections at redshifts
$z=0.011$ and $z=0.027$. \citet{nic05} reported the WHIM absorption from
transitions of   
\neix\ K$\alpha$, \ovii\ K$\beta$, \oviii\ K$\alpha$, \ovii\ K$\beta$, 
\nvii\ K$\alpha$, and \civ\ K$\alpha$ (Table~2 and Figure~8 in
\citealt{nic05}), whose rest-frame wavelengths are 13.447, 18.629, 18.969,
21.602, 28.787, and 33.736 \AA, respectively \citep{ver96}. 
Therefore, we focus our
attention on spectral ranges covering these transitions.

Obtaining a good continuum model is crucial to absorption-line
measurement. Since all coadded \chandra\ spectra
contain the contribution from HRC observations, to account for
grating order-overlapping issues (Section~\ref{sec:obs}), we decide to fit
a broad range of the spectra from 7 to 40 \AA, which covers the transitions
to be examined. For a ease of comparison of three \chandra\ spectra, we
jointly fit the Spectra I, II, and III. We first use a
Galactic-absorption modified power-law ({\sl wabs*powerlaw} in xspec) 
to fit the
continuum emission, with neutral hydrogen column density $N_{\rm H}$
linked together
while the power-law index $\Gamma$ and normalization are 
allowed to vary.
We obtain an unacceptable fit with $\chi^2=11646$ over 3590 degrees of
freedom (DOF), and we find that some ``broad'' features cannot be
accounted for by this simple model. These features are
mainly due to imperfect instrumental calibration around node boundaries,
CCD chip gaps plus dithering effects, the oxygen absorption edge, and
imperfect cross calibration between ACIS and HRC observations 
\citep{mar04, nic05}. We then use Gaussian profiles to compensate the
uncounted calibration residuals. To minimize the effect of the known strong
ISM absorption 
lines on continuum modeling, we include narrow absorption lines 
(with widths $\sigma\lsim50~{\rm km~s^{-1}}$) of \oi, \oii,
\cvi, \ovi, \ovii, and \neix\ K$\alpha$ and \ovii\ K$\beta$ transitions 
at their rest-frame wavelengths \citep{ver96, jue06, yao09}
in our spectral fitting. The centroid
energies and widths of these Gaussians 
are linked together while the normalizations are
allowed to vary in the jointly spectral fitting. We find that, besides
these narrow absorption lines produced in the ISM, we need 10 broad
and one narrow Gaussian profiles to obtain an acceptable fit, and
locations of these Gaussian are 10.53, 13.80, 14.60, 17.90, 18.19, 19.02, 
19.10, 23.24, 23.31, 23.66, and 29.89 \AA\ with widths ($\sigma$) of 
0.61, 2.56, 0.04, 0.00054, 0.17, 1.36, 0.10, 1.45, 1.69, 0.41, and 
0.39 \AA. The final $\chi^2=4448.2$ with 3512 DOF and the constrained
$N_{\rm H}=1.092\pm0.06\times10^{20}~{\rm cm^{-2}}$, 
and $\Gamma=2.131\pm0.001$, $2.011\pm0.001$, $2.067\pm0.001$ for Spectra I,
II, and III, respectively. 
\emph{Because the absorption line measurement depends
only on the \emph{local} continuum, and because the centroids of these
Gaussian profiles do not directly superpose on the wavelengths of the
putative WHIM absorption lines, these Gaussian profiles will not have any
effect on the line equivalent width measurement conducted below.}
\xmm\ spectrum (Spectrum IV) does not have the
grating-order-overlapping issue, so we use the {\sl wabs*powerlaw} model
to fit the six narrow spectral ranges as plotted in Figure~\ref{fig:rgs}. We
exclude the bad pixels in our spectral analysis, and
obtain $\chi^2/DOF=1.36, 1.49, 1.49, 1.28, 1.75, 1.56$ with DOF of
137, 75, 103, 105, 116 and 104, respectively. 
Figures~\ref{fig:chan-Ni}-\ref{fig:rgs}
show these best fit continua with the ISM absorption lines removed from
spectral models.

Now, let us examine the existence of the reported WHIM absorption
lines. Figures~\ref{fig:chan-Ni}-\ref{fig:rgs} show the same portions of
\chandra\ and \xmm\
spectra as shown in Figure~8 of \citet{nic05}, in which they demonstrated
their detections of the WHIM absorption lines at $z=0.011$ and $z=0.027$
from transitions of \neix\ K$\alpha$, \ovii\ K$\beta$, \oviii\ K$\alpha$,
\ovii\ K$\beta$,  \nvii\ K$\alpha$, and \civ\ K$\alpha$. In particular,
Figure~\ref{fig:chan-Ni} reproduces their Figure~8, as they were extracted
from the same \chandra\ observations. 
Several bad pixels of the RGS accidentally fall around the 21.8 \AA\ region 
and cause an instrument feature (Figure~\ref{fig:rgs}). Although a 
careful calibration could yield a correct instrumental response of 
the RGS, in this work we do not use the \xmm\ spectrum to assess the WHIM 
detection at $z=0.011$.  Visual inspection reveals
that none of the above WHIM lines reported by \citet{nic05} is 
detected in the previously \chandra\ observations they analyzed
(Figure~\ref{fig:chan-Ni}), and in the newly available \chandra\ observations
(Figure~\ref{fig:chan-new}), and in the \xmm\ observations
(Figure~\ref{fig:rgs}).

The reported WHIM detections could be due to the improper continuum
placement on the spectrum. Visual inspection also reveals that there might
be an absorption feature at $\sim21.8$ \AA\ in Spectrum I, which
corresponds to the 
reported \ovii\ K$\alpha$ WHIM absorption line at $z=0.011$
(panel (c) in Figure~\ref{fig:chan-Ni}).
However, such an absorption feature is not visible in
Spectra II and III (panel (c) in Figures~\ref{fig:chan-new} and
\ref{fig:chan-all}), 
and its significance in Spectrum I 
depends largely on how the local continuum is placed. Please note that our 
spectral
profiles of the local continua were obtained by jointly fitting the 
Spectra I,
II, and III (see above), and the best-fit model seems to account for the
general spectral variation reasonably well. To quantitatively assess the
existence of the \ovii\ K$\alpha$ WHIM line at $z=0.011$, in the joint
analysis of the Spectra I, II, and III, in addition to the 10 Gaussian
profiles, we add the 11th ``broad'' Gaussian to the local continuum and 
another narrower Gaussian with width
\footnote{Different widths yield nearly identical results as long as the
  line is still unresolved by the instrument. Please see
  Section~\ref{sec:exp} for justifications for the line width.}
$\sigma=50~{\rm km~s^{-1}}$ 
to represent the WHIM absorption line. Again, we link the line centroids
and widths together in our joint fitting but allow the normalizations to
  vary among the Spectra I, II, and III. The best fit yields a line
  centroid of 21.90 \AA\ and $\sigma=0.050$ \AA\ for the ``broad'' 
component and a centroid of 21.87 \AA\ for the narrow component. 
The red curves in panel (c) of 
Figures~\ref{fig:chan-Ni}-\ref{fig:chan-all} show
  the modified local continua. These additional
  components reduce the $\chi^2$ of 26.4 by adding nine DOF in total. 
However, applying the best-fit models to individual spectra yields
a $\chi^2$ change of 13.7, 4.3, and 10.4 by introducing  
  five additional DOF to spectral fitting of Spectrum I, II, and III, 
respectively. All these \chandra\ spectra can be well described 
by the same continuum profile with various normalizations, which has been
proved to be reasonable to other wavelength ranges (panels (a)-(b) 
and (d)-(f) in Figures~\ref{fig:chan-Ni}-\ref{fig:chan-all}). Because we find
essentially no improvement in fitting the Spectrum II,
adding the ``broad'' Gaussian profile to the local 
continuum around 21.9 \AA\ is statistically unnessary. 

We next attempt to measure the equivalent widths (EW) of the putative WHIM
absorption lines. Since we have not consistently detected any WHIM absorption 
of any transitions reported by \citet{nic05}, we focus here on measuring the 
upper limit to the equivalent width (EW) of the 
\ovii\ K$\alpha$ line, which is expected to be the most abundant transition 
in a broad temperature range of the WHIM (Section~\ref{sec:intro}). 
At redshifts $z=0.011$ and $z=0.027$, the rest-frame \ovii\ K$\alpha$ at 
21.60 \AA\ is shifted to 21.84 and 22.19 \AA, respectively, and therefore our
measurements are obtained from the spectral range 
described by the continuum colored as blue in panel (c) of 
Figures~\ref{fig:chan-Ni}-\ref{fig:rgs}. To make a fair comparison with the 
results obtained by \citet{nic05}, we also make similar measurements by using 
the continuum colored as red. We should emphasize that the latter measurements
are only for comparison purposes, since the additional component to the 
local continuum that ``amplifies'' the significance of the absorption line 
at 21.8 \AA\ is statistically unnecessary. Our results are reported in 
Table~\ref{tab:results}. 

In summary, our analysis and results do not support the existence of the 
two WHIM systems at $z=0.011$ and $z=0.027$. We analyze all the available 
\chandra\ and \xmm\ observations of Mrk~421, and have not seen any consistent
WHIM absorption of any transition at either $z=0.011$ or $z=0.027$ reported
by \citet{nic05}. We measure the EW of the \ovii\ 
K$\alpha$ absorption line. For the system at $z=0.027$, we obtain a firm 
3$\sigma$ upper limit of $EW<1.5$ m\AA\ from the spectrum (Spectrum II)
extracted from the 
newly available \chandra\ observations, in contrast to $EW=2.2\pm0.8$ m\AA\ 
obtained by Nicastro et al. For the system at $z=0.011$, we confirm the 
existence of a small dip in the spectrum (Spectrum I)
identical to that used by \citet{nic05}, 
but the similar spectral feature is absent in
Spectrum II. The measurement of the line heavily depends on how 
the local continuum is placed, and the additional spectral component 
that amplifies its significance is statistically unnecessary. 
We obtain an upper limit of $EW<1.5$ m\AA\ from the new spectrum,
in contrast to $EW = 3.0^{+0.9}_{-0.8}$ m\AA\ reported  
 by \citet{nic05}. The findings
in this work are consistent with the previous 
investigation made by \citet{kaa06}, in which the authors scrutinized the
same data set used by Nicastro \etal\ and a subset of calibration
observations of
\chandra\ and \xmm. Therefore, we conclude that 
there is no WHIM line detected at either $z=0.011$ or $z=0.027$ 
along the Mrk~421 sight line.

There are also other important absorption lines in the spectral
range of interest (Figures~\ref{fig:chan-Ni}-\ref{fig:rgs}). 
The prominent lines are \ovii, \neix, and \civ\ K$\alpha$ 
and \ovii\ K$\beta$ at $z\approx0$, and the \ovi\ K$\alpha$ with 
rest-frame wavelength 22.0 \AA\ at $z\approx0$ 
is also visible in all four spectra albeit less significant. Again, 
we only measure the EWs of \ovii\ K$\alpha$  
and report them in Table~\ref{tab:results}. The \ovii\ line at $z\approx0$ can
be well explained as absorption from 
the Galactic diffuse interstellar medium (ISM)
\citep{yao07}, and the \ovi\ line is also believed to originate from the ISM
\citep{sav05}. 
It is worth noting that the EW
of \ovii\ K$\alpha$ at $z\approx0$ measured from \xmm\ observations is 
$\gsim3\sigma$ larger  
\footnote{The differential significance is calculated as 
\[|EW_1 - EW_2|/\sqrt{(\Delta EW_1)^2 + (\Delta EW_2)^2}, \]
where $\Delta EW$ is the $1\sigma$ error of $EW$ measurement.}
than that obtained from \chandra\ observations
(Table~\ref{tab:results}). 
Such a discrepancy in \ovii\ has also been measured by \citet{kaa06}. 
The causes to this discrepancy are still under investigation, and the
results will be published elsewhere. 
In this paper, we mainly focus on X-ray absorption measurements of 
the WHIM and will not discuss these non-WHIM measurements any further. 

\section{Bootstrap Simulations}
\label{sec:sim}

In Sections~\ref{sec:obs} and \ref{sec:res}, we reprocessed 
all the available
observations and obtained two spectra with the highest S/Ns ever
collected with \chandra\ and \xmm. Despite the increased S/N, we still
cannot confirm the X-ray WHIM detections reported by \citet{nic05}. Given
such a frustrating situation (see also discussions in
Section~\ref{sec:intro}),  
several crucial questions need to be answered before further major time
allocations search for the WHIM through X-ray 
absorption line measurement: How high S/Ns are needed to firmly establish a
WHIM detection? What is the \ovii\ column density range suitable for
\chandra\ and \xmm\ to explore with reasonably long exposure time? Is it
possible to conduct a systematic study of the WHIM using the current X-ray
telescopes? If not, what are the requirements for next-generation X-ray 
telescopes?

\subsection{Expected WHIM properties and required instrumental sensitivities}
\label{sec:exp}

\begin{figure}
  \plotone{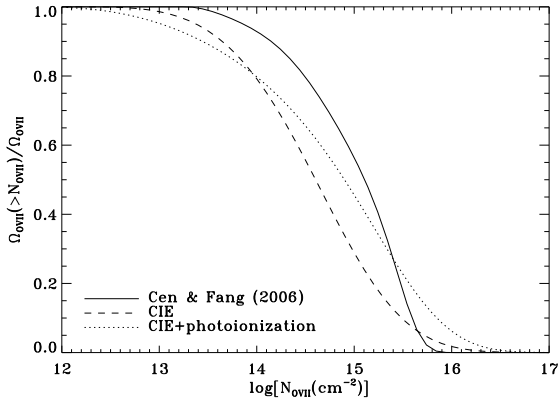}
  \caption{Fraction of baryons contained in the \ovii-bearing gas as a
  function of \ovii\ column density. The solid line is derived from
  Figure~5 in \citet{cen06a}. The dashed and dotted lines are derived
  from figures in \citet{smi12}, which consider 
  collisional ionization (CIE) only and CIE plus photoionization,
  respectively. See text for details.
  }
  \label{fig:fang}
\end{figure}

Before running bootstrap simulations to address these questions, we first
review the WHIM properties based on the results of numerical simulations. 
We then estimate the required instrumental sensitivity to detect the WHIM
by answering the following two specific questions: First,
what \ovii\ column density is needed to probe the majority of the baryons
contained in the \ovii-bearing WHIM? Cosmological simulations predict
different \ovii\ column density distributions per unit redshift with
different recipes for galactic feedback 
and intergalactic photoionization \citep{fang02b, cen06a, smi12}. In
Figure~\ref{fig:fang} we convert two recent distributions (Figure~5 in
\citealt{cen06a} and figures in \citealt{smi12}) 
to the fraction of baryons contained in the \ovii-bearing WHIM as a
function of \ovii\ column density.  
In this conversion, we assume a single temperature and uniformly
distributed metallicity for all absorbers. However, \citet{smi12} 
find covariance between $N_{\rm OVII}$ and metallicity, and the
WHIM is expected to be multiphase. To obtain a more realistic
distribution, one should consider the ionization and metallicity
correction together. However, without a detailed description of the WHIM
properties (which in fact are our final goals), such a correction is
difficult to apply. Considering the general trend that higher column
density absorbers have higher metallicity, the plotted baryon fractions
contained in the higher column absorbers (e.g., 
$N_{\rm OVII}\gsim10^{15}~{\rm cm^{-2}}$) should be regarded as upper
limits. Furthermore, since there is no clear sign of convergence at low
column densities
\citep{cen06a, smi12}, in our conversion we use a 4- or 6-degree
polynomial to fit the number density distributions, which artificially
generate turnover at column densities 
$10^{10}-10^{12}~{\rm cm^{-2}}$. The 
low column density absorbers could contain even more baryons if the
distributions converge at lower columns. Nevertheless,
Figure~\ref{fig:fang} gives an estimate of the baryon fraction
distribution, which indicates that $\sim$55\% (or 65\% or 80\%,
depending on which simulation is selected) of baryons are contained in
absorbers with $N_{\rm OVII}>10^{14.5}~{\rm cm^{-2}}$ and that $\sim$30\%
(or 45\% or 55\%) of baryons are contained in absorbers with 
$N_{\rm OVII}>10^{15}~{\rm cm^{-2}}$.

Another important, yet to be determined property is the dispersion/Doppler
velocity $b_v$ 
of the \ovii-bearing gas. This property, although not essential 
for the detectability of \chandra\ and \xmm\
(Section~\ref{sec:sim-chan-xmm}), is important for estimating the
required spectral resolution of future X-ray instruments
(Section~\ref{sec:ixo}). Here we use \ovi\ 
as a reference. Assuming most detected IGM \ovi\ tracing the WHIM
at temperatures $\sim10^{5.5}$~K and the total $b_v$ being a combination of
thermal and turbulent motions through quadrature
$b_v^2 = b_{\rm th}^2 + b_{\rm turb}^2$, the median  
$b_v\sim30~{\rm km~s^{-1}}$ of the \ovi\ absorption lines 
\citep{dan08,tri08} suggests a turbulent velocity of
$b_{\rm turb}\sim24~{\rm km~s^{-1}}$ for the \ovi-bearing gas. Higher
temperature gas may have higher turbulent velocity. Thus $b_v$ of
the \ovii-bearing gas at temperatures $\sim10^6$~K is expected to be
$>40~{\rm km~s^{-1}}$. Another useful reference is the hot phase Galactic
ISM traced by the \ovii\ absorption line at $z\approx0$, although the IGM
environment could be substantially different from the ISM. Nevertheless,
along the Mrk~421 and LMC~X-3 sight lines, the $b_v$ of the
\ovii\ gas can be as large as $100~{\rm km~s^{-1}}$ 
\citep{wang05, yao07, yao09}. We therefore will explore three $b_v$ values, 
50, 75, and 100~${\rm km~s^{-1}}$, in our simulations. 

Second,
what is the flux limit that X-ray telescopes should reach in
order to statistically investigate the WHIM? The answer to this question
depends largely on how many WHIM systems are planned to detect. Clearly
when more systems are detected, one can investigate WHIM properties, like
thermal and dynamic properties, metallicities, and their evolution with
redshift. To obtain a spectrum with a requested S/N, the required
exposure time is inversely proportional to the flux of a QSO. And the
higher the redshift of a QSO, the higher possibility there is
intervening WHIM 
absorbers along the line of sight. Therefore, to estimate the number of
WHIM systems that could be detected, one should create a target list by
considering the QSO redshift and flux as a whole. 

In this work, instead of
providing a target list for the WHIM investigation, we require that future
X-ray telescopes be able to observe and search for the 
corresponding \ovii\
absorbers along the majority of the sight lines where IGM \ovi\
absorbers have already 
been reported (Section~\ref{sec:intro}). In this regard, we
examine the  
target lists in \citet{dan08} and \citet{tri08} and find that there are
$\sim80$ IGM \ovi\ systems along 28 QSO sight lines reported so far. 
Among them, we have assembled the available soft X-ray fluxes 
\footnote{We use a power-law form $f(E)\sim E^{-1.1}dE$ to convert the 
reported flux to the energy range of 0.5-2.0 keV (Section~\ref{sec:intro}).}
for 24 sight lines  
from the bright source catalogs of {\sl Einstein}, {\sl ASCA}, 
{\sl BeppoSAX}, and {\sl ROSAT}, and from several pointing observations with 
\xmm\ and \chandra\ \citep{wil87, ueda05, ree00, ver07, bia09, pap07,
  fos06, por04, whi00, lei07}. These fluxes range
from 0.006 to 46.5 mCrab
with a median of $\sim0.3$ mCrab. Fifteen of them have fluxes $\gsim0.2$
mCrab and sample a total redshift pathlength of $\Delta z=2.1$. 
For the WHIM \ovii-absorbers with $N_{\rm OVII}>10^{15}~{\rm cm^{-2}}$,
\citet{cen06} predicted the absorber frequency to be  
$d{\cal N}/{dz}\sim7$, and \citet{smi12} predicted 
the frequency to be $\sim20$ considering collisional
ionization only and $d{\cal N}/{dz}\sim5$ if 
photoionization is included. For absorbers with 
$N_{\rm OVII}>10^{14.5}~{\rm cm^{-2}}$, $d{\cal N}/{dz}$ is predicted to
be $\sim18$, 12, and 110, respectively.
Therefore, X-ray telescopes should be able to facilitate absorption
line study from background sources with flux of 
$\sim0.2$ mCrab, and $\ge$10 
WHIM systems with
$N_{\rm OVII}>10^{15}~{\rm cm^{-2}}$ and $\ge$24 systems with
$N_{\rm OVII}>10^{14.5}~{\rm cm^{-2}}$ are expected in these sources
along which the corresponding \ovi\ absorptions are already known.

\subsection{Detection limits of current X-ray telescopes}
\label{sec:sim-chan-xmm}

\begin{figure}
  \plotone{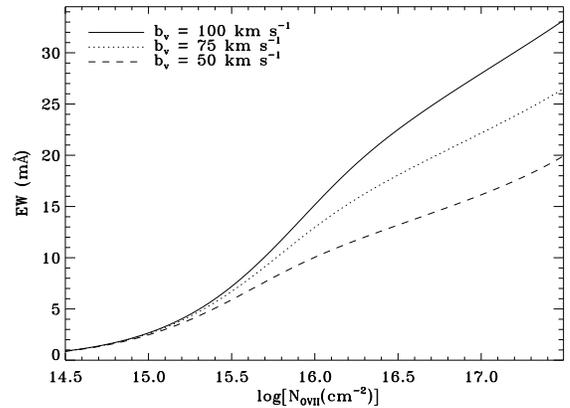}
  \caption{Curve of growth of the equivalent width of \ovii\ K$\alpha$ 
  absorption line versus \ovii\ column density with three different Doppler 
  dispersion velocities ($b_v$).
  }
  \label{fig:cog}   
\end{figure}

We now use bootstrap simulations to examine the detectability of 
\chandra-LETG and
\xmm-RGS. In our simulations, we utilize the user-developed 
absorption line model {\sl absline} \citep{yao05} to model \ovii\ absorbers
with column density $N_{\rm OVII}$ and dispersion velocity $b_{v}$, and use
a PL continuum to model an emission spectrum. With the input 
parameters of $N_{\rm OVII}$, $b_{v}$, PL normalization, and an
exposure time, we simulate two spectra within Xspec, using the RMFs and ARFs
produced in Section~\ref{sec:obs}, to mimic \chandra\ observations 
and \xmm\ observations, respectively. From the simulated spectra, we use  
a Gaussian profile to represent the absorption line and measure its
EW and uncertainty $\Delta EW$ at the expected wavelength position and then
calculate the line SL (Eq.~\ref{equ:sig}). For 
a set of parameters, we repeat this procedure 1000 times and record the 
measurement SL of the line. We then sort the SL and select the 100th
value (i.e., 90\% simulation trials yield the SL equal to or higher than the 
selected value). 
We make the same measurement for various combinations of different exposure
times, $b_{v}$, and $N_{\rm OVII}$. Since the full-width-half-maximum
(FWHM) of the LSF of both \chandra-LETG and \xmm-RGS is
$\sim$50~m\AA, which is much broader than the expected 
$b_v$ values of the WHIM \ovii\ absorbers (Section~\ref{sec:exp}), all
simulated absorption lines appear unresolved. What really matters for
measurement is the effective EW of an absorption line. To provide a direct
comparison to observations, we convert the combination of 
$N_{\rm OVII}$ and $b_{\rm v}$ to EW through the curve-of-growth (COG), and
the combination of exposure time, 
RMF, and ARF to S/N per 50-m\AA\ resolution
element.  For reference, we plot the COG of
\ovii\ K$\alpha$ line in Figure~\ref{fig:cog}.

\begin{figure*} 
  \plotone{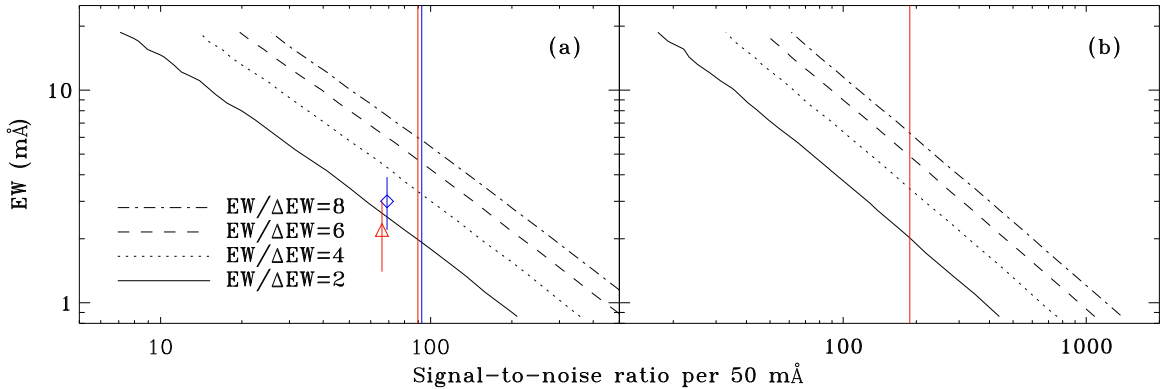}
  \caption{The 2, 4, 6, and 8$\sigma$ detection significance of 
    \ovii\ K$\alpha$ versus
    S/N per 50-m\AA\ resolution element with \chandra-LETG (a) 
    and \xmm-RGS (b). The
    vertical red and blue lines mark the S/N levels of the final coadded
    spectra around the clamied 
    \ovii\ lines at $z=0.027$ and $z=0.011$,
    respectively. The red triangle and blue diamond indicate the
    detections and their 1$\sigma$ ranges reported by \citet{nic05}.
    In simulations, the instrumental response files were taken from \chandra\
    observation with ID 4148 and \xmm\ observation with ID 0099280201.
} 
  \label{fig:cha-xmm}
\end{figure*}

Figure~\ref{fig:cha-xmm} shows our simulation results. Taking the simulations
for \chandra-LETG for an example, our results can be interpreted as:
from a spectrum with S/N $=50$ per resolution element, an \ovii\
K$\alpha$ absorption line with EW $=3.4$, 5.7, 8.1, 10.3~m\AA\ can be
measured at $\gsim2$, 4, 6 and 8$\sigma$ SLs, respectively, in 900 out of
1000 observations. In another word, to
measure the absorption line with these EWs at 2, 4, 6 and 8$\sigma$ or
higher SLs, a spectrum with S/N $\ge50$ is required.
Interestingly, the
measured line significances behave nicely like PL functions
 in the EW-S/N space. For reference, we formulize them as,
\begin{equation}
\label{equ:sig}
\log(EW) = \alpha \times \log(S/N) + \sum\limits_{i=0}^3\beta_i\times SL^i,
\end{equation}
where $SL$ is 2, 3, 4, 5, 6, 7, or 8.
For \chandra\ simulations, $\alpha=-0.932$, $\beta_{0-3}=1.79108$, 
0.188439, $-$0.0148031, and 0.000476159, and for \xmm, $\alpha=-0.970$, 
$\beta_{0-3}=2.14183$, 0.217703, $-0.0197475$, and 0.000748764.

Let us use our simulation results to further examine the two putative 
WHIM systems. In Figure~\ref{fig:cha-xmm}, we also plot the EWs 
reported by \citet{nic05} and their 1$\sigma$ errors
at the corresponding S/N positions of Spectrum I. We find that their 
reported SLs appear to be consistent with expectations of
our simulations.
However, we stress again that their SL measurement could have been
largely impacted by the improper continuum placement
(Section~\ref{sec:res}). Our simulations indicate that an 
\ovii\ K$\alpha$ absorption line with EW=3.0 m\AA\ 
at $z=0.011$ \citep{nic05} should be measured at $\ge3.7\sigma$ SL from 
\chandra\ Spectrum III with S/N$\sim92$ 
(Eq.~\ref{equ:sig}; Figure~\ref{fig:cha-xmm}).
This is in contrast to the fact that we only obtain an upper limit
(Figure~\ref{fig:chan-all}; Table~\ref{tab:results}).  
At $z=0.027$, an \ovii\ line with EW $=2.2$ m\AA\ should be detected at
$\gsim2.3\sigma$ from the coadded \chandra\ and \xmm\ spectra, which,
again, is in contrast to the fact that we only obtain upper limits. 
We therefore conclude that the putative \ovii\ WHIM absorptions, 
if they exist at all, must have lower EWs than reported. 

Is the reported WHIM absorption at $z=0.027$ associated with a galaxy
filament?  \citet{wil10} discovered a filament of late-type galaxies 
at $z=0.027$, suggesting a 
possible association between the reported WHIM absorber and the filament.
In the spectrum of PKS~0405-123 observed with the Cosmic Origins
Spectrograph (COS) aboard the {\sl Hubble Space Telescope}, a similar
association has also been suggested by a broad \ovi\ 
absorption line \citep{sav10}. The authors estimated that the WHIM
contained within the abundant late-type galaxy groups could make up to
15\% of the total baryons in the local 
universe. However, the non-detection of \ovii\ at $z=0.027$ in this work
leads us to conclude that there is, if at all, much less \ovii\ 
than reported. A recent search in a high-S/N COS
spectrum of Mrk~421 also failed to find any absorption
feature of BLA at the redshift \citep{dan11}.  
Similarly, there is no detected hot intragroup 
\ovii-bearing gas in our Local Group  
\citep{fang06, yao07, bre07}.  Therefore, if there is any gas associated
with the galaxy filament at $z=0.027$ along the Mrk~421 sight line, 
it must be at higher temperatures ($T>10^{6.5}$ K) and/or be metal poor
and thus contain too little \ovii\ and \hi\ to be detected.

Our simulation results also have important implications for using 
current X-ray observatories to search for the WHIM absorption.
First, \chandra\ and \xmm\ can probe the WHIM absorbers with 
$N_{\rm OVII}\gsim10^{16}~{\rm cm^{-2}}$ with exposure times 
of $\sim1$ Ms from a QSO spectrum with a flux of $\sim0.2$ mCrab. 
The \ovii\ column densities correspond to EWs of 
$\gsim10-15$ m\AA\ (Figure~\ref{fig:cog}), which require a \chandra\ 
(\xmm) spectrum with S/N $=18-27~(41-63)$ to measure it at $\gsim4\sigma$ 
significance levels (Figure~\ref{fig:cha-xmm}; Eq.~\ref{equ:sig}). To
collect a spectrum with such S/N, takes 1.4--3.1 Ms for
\chandra\ (adopting $A=20~{\rm cm^2}$ effective area) and 3.6--8.5 Ms for
\xmm\ (adopting $A=40~{\rm cm^2}$ effective area) from a QSO 
with 0.2 mCrab flux. This is consistent with the commonly detected \ovii\
absorption lines, which trace the Galactic hot ISM, in
the spectra of local QSOs ($z\lsim0.1$) with fluxes $>2$ mCrab
\citep{fang06,bre07,yao07}. However, the simulated $d{\cal N}/dz$
of \ovii\ drops by
about two orders of magnitude between $N_{\rm OVII}>10^{15}~{\rm cm^{-2}}$ 
and
$N_{\rm OVII}>10^{16}~{\rm cm^{-2}}$ \citep{smi12}, meaning that
such high column \ovii-absorbers are extremely rare and may only exist in
the circumgalactic environment around galaxy structures 
(e.g., \citealt{fang10}).

Second, it is likely infeasible to conduct a systematic study
of the WHIM with \chandra\ or \xmm. The EW for 
$N_{\rm OVII}=10^{15}~{\rm cm^{-2}}$ is $\sim2.5$ m\AA\
(Figure~\ref{fig:cog}),  which requires a spectrum with S/N$\sim120$ 
(\chandra) and 260 (\xmm) to measure it at $\sim4\sigma$ significance
level (Figure~\ref{fig:cha-xmm}). It would require 61 Ms for 
\chandra\  and 144 Ms for \xmm\ to
collect a spectrum with such a high S/N from a QSO with 0.2 mCrab
flux. Even for a blazar at its burst state like Mrk~421, 
it still takes 0.6 (1.4) Ms \chandra\ (\xmm) time from the
source continuously being bright at 50 mCrab level. It would take 341 Ms
for \chandra\ and 804 Ms for \xmm\ to complete the
survey of the 15 background QSOs with fluxes $\gsim0.2$ mCrab along which 
IGM \ovi\ has been reported (Section~\ref{sec:exp}).
Furthermore, since the
expected WHIM \ovii\ absorption lines cannot be resolved,
it is impossible to study the dynamics of the WHIM
with either \chandra-LETG or \xmm-RGS. Clearly, we need new 
X-ray spectrographs with much improved 
spectral resolution and much larger collecting area to 
systematically study the WHIM.

\subsection{Requirement for next generation X-ray telescopes}
\label{sec:ixo}

\begin{figure*}
  \plotone{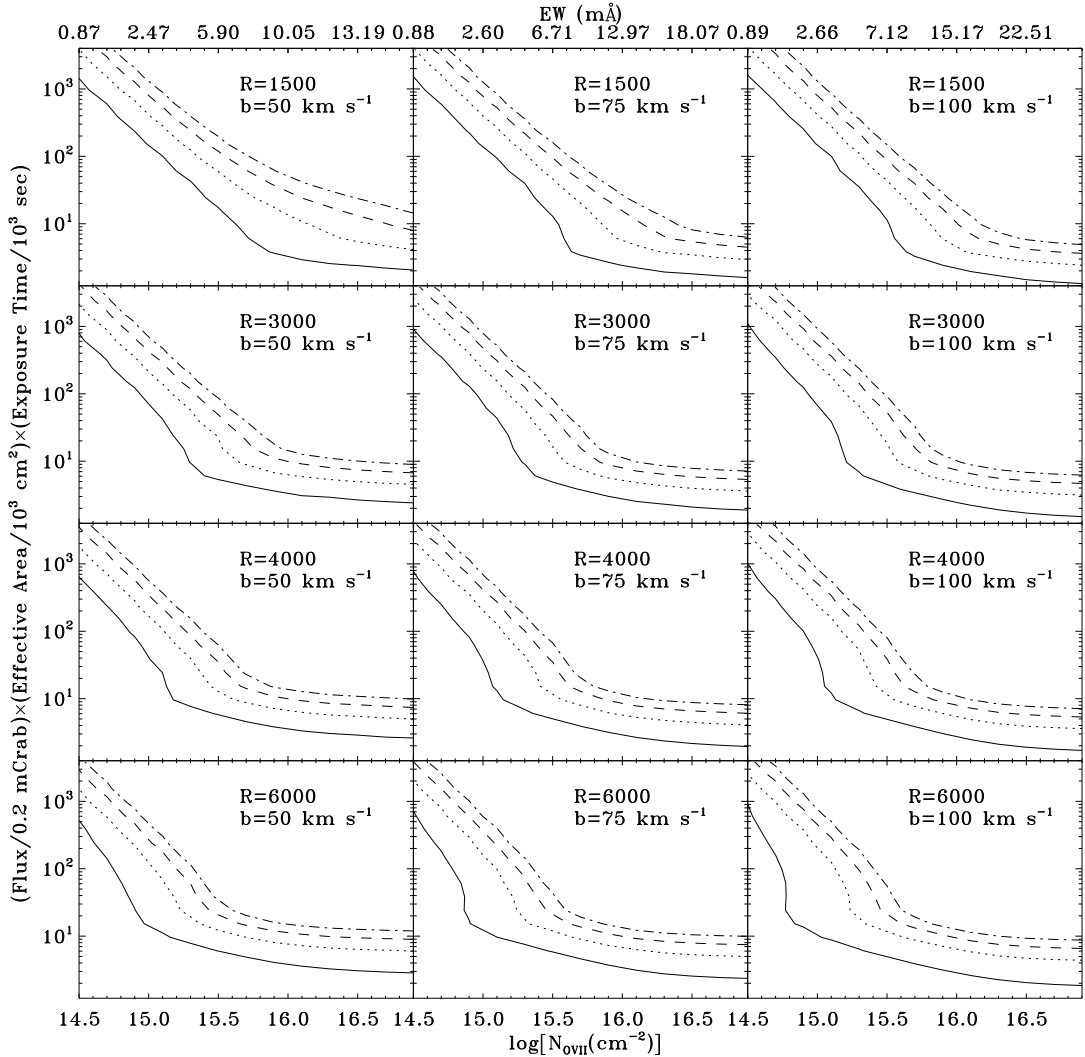}
  \caption{The required exposures for detecting \ovii\ K$\alpha$ line at 2,
    4, 6 and 8$\sigma$ significance levels (corresponding to curves from 
    bottom to top in each panel) for the absorbing gas with dispersion
    velocities of 50, 75, and 100~${\rm km~s^{-1}}$ (panels from left to
    right) and column densities 
    $\log N_{\rm OVII} ({\rm cm^{-2}})=14.5-17.0$. The top X-axis shows
    the corresponding equivalent width in m\AA.
    The simulations used flux integrated from 0.5-2 keV and assumed a 
    Gaussian profile of the line spread function of
    a spectrograph with spectral resolution  
    $R\equiv\lambda/$FWHM $=1500$, 3000, 4000, 6000 (panels from 
    top to bottom),
    respectively. 
  }
  \label{fig:ixo}
\end{figure*}

The goal in this section is to establish the required spectral
resolution and effective area for next 
generation X-ray telescopes. Since there is no existing X-ray telescope with
spectral resolution better than those of \chandra\ and \xmm,
we use Gaussian profiles to simulate the LSFs of the putative X-ray
spectrographs. We assume a spectral resolution, which is defined 
\footnote{We use the FWHM of the LSF as $\Delta\lambda$.}
as $R=\lambda/\Delta\lambda$,  
of 1500, 3000, 4000, and 6000,
respectively, with effective area of $A=1000~{\rm cm^2}$ 
throughout the whole wavelength coverage. 
We use four spectral bins within a FWHM spectral range (i.e., 
an oversample factor of 4) in our simulations.
With a set of selected $R$, $N_{\rm OVII}$, and $b_v$ of the
\ovii-bearing gas, we follow the same procedure as
described in Section~\ref{sec:sim-chan-xmm} to simulate an absorption spectrum 
of an background source with 0.2 mCrab flux for an assumed exposure time,
and then measure the EW and $\Delta$EW from the simulated spectrum. 
We repeat the procedure 1000 times and select the SL value so that 90\% trials
have the same or higher SL. In a real observation, background source flux, 
instrumental effective area, and exposure time can compensate
each other, so we define their product (FAE) as a figure of
merit. Figure~\ref{fig:ixo} shows our simulation results. For 
reference, we also label the corresponding EW to the labeled 
$N_{\rm OVII}$ at top X-axis of the plots. 
Table~\ref{tab:ixo} summarizes the required exposure time for the
future spectrograph with 1000~cm$^2$ effective area and different spectral
resolutions to constrain the \ovii\ absorption with 
$N_{\rm OVII}=10^{15}~{\rm cm^{-2}}$ and three different $b_v$ values
at $>4$ and 6$\sigma$ significance levels from a background source
with 0.2 mCrab flux. 

\begin{deluxetable}{c|ccc}
\tablewidth{0pt}
\tablecaption{ \label{tab:ixo}}
\tablehead{
& \multicolumn{3}{c}{Dispersion velocity $b_v$} \\
$R$ & 50 km~s$^{-1}$ & 75 km~s$^{-1}$ & 100 km~s$^{-1}$} 
\startdata
1500 & 416.9 (808.8) & 398.8 (794.6) & 414.6 (799.5)\\
3000 & 217.4 (433.9) & 237.6 (492.6) & 277.8 (549.9) \\
4000 & 171.1 (350.3) & 212.1 (425.7) & 241.2 (523.0) \\
6000 & 126.3 (276.5) & 162.9 (368.2) & 208.3 (461.7)
\enddata
\tablecomments{Required exposure times in unit of ks 
for future X-ray spectrographs with various spectral resolutions to
obtain $>4\sigma$ significance measurement of 
\ovii\ K$\alpha$ with $N_{\rm OVII}=10^{15}~{\rm cm^{-2}}$ 
and different
dispersion velocities. The values in parentheses are required
exposures for $>6\sigma$ constraint (Figure~\ref{fig:ixo}). 
}
\end{deluxetable}

Our simulations results deserve some explanation. First, each significance
contour has a steep PL plus a relatively flat floor in the logarithm  
FAE-$N_{\rm OVII}$ space, which roughly corresponds to the shape of the COG
(Figure~\ref{fig:cog}). The transition locations indicate the column
densities from which the saturation of a line begins to affect the
measurement. Second, to constrain 
$N_{\rm OVII}=10^{15}~{\rm cm^{-2}}$ at the same significance levels, it
takes an increasingly long time with increasing $b_v$ for all 
resolutions except for $R=1500$ (Table~\ref{tab:ixo}). This is mainly due
to line broadening. With $R=1500$, no lines with $b_v<120~{\rm km~s^{-1}}$
can be resolved, since the intrinsic line width is narrower than the
instrumental LSF. Therefore the simulations are analogous to those for
\chandra\ and \xmm, which explains the required exposure times being  
essentially same for all $b_v$ values. At $R\gsim3600$, a line with 
$b_v>50~{\rm km~s^{-1}}$ becomes resolved. Because the column density  
$N_{\rm OVII}=10^{15}~{\rm cm^{-2}}$ lies on the linear part of the COG 
(Figure~\ref{fig:cog}), an increased $b_v$ does not increase the EW but
broadens the line profile, and therefore requires longer exposure to pick
up the absorption signals from the continuum. 

A spectrograph  with spectral resolution of $R\gsim3600$ is required
to resolve the expected WHIM \ovii\ absorption lines. As we discussed
in Section~\ref{sec:exp}, the lower 
limit to the dispersion velocity of the WHIM \ovii\ absorbers is expected
to be $b_v\gsim50~{\rm km~s^{-1}}$, which is equivalent to a FWHM of 
$83.3~{\rm km~s^{-1}}$ for a Gaussian profile fit. Although we can measure
$b_v$ by comparing the different saturation levels of the
K$\alpha$ and K$\beta$ transitions of \ovii\ from the ``low'' resolution
observations made with \chandra\ and \xmm\ (e.g., \citealt{yao05, yao07}),
such technique cannot be applied to a column density as low as 
$N_{\rm OVII}\sim10^{15}~{\rm cm^{-2}}$ because saturation is
negligible and the K$\beta$ line becomes too weak to be detectable. 
Therefore, we have to rely on the instrumental resolution 
($R\sim3600$) to resolve the line and obtain the dynamical and kinematic 
information of \ovii-absorbers. Please note that, to date, the X-ray 
spectrograph with the highest spectral resolution is the High Energy 
Transmission Grating Spectrograph (HETGS; \citealt{can86}) aboard 
on \chandra, which has two sets of gratings: the high energy grating (HEG) 
and medium energy grating (MEG) with resolutions of $\sim1700$ and 
$900$ at 20 \AA. Clearly, they are still not high enough for resolving the 
expected WHIM absorption line. Furthermore, the small effective area 
($<1~{\rm cm^2}$ and $\sim2~{\rm cm^2}$) around 20 \AA\ makes the HETGS 
infeasible for the task discussed here (please also see below).
Figure~\ref{fig:lsf} shows the necessity for high 
spectral resolution by
comparing the LSFs of \xmm, \chandra, and a
next-generation X-ray spectrograph with $R=4000$ and the 
corresponding simulated spectra. It demonstrates
that an absorber with two velocity components separated by 
$150~{\rm km~s^{-1}}$ cannot be resolved by \xmm\
(\chandra) even from a spectrum with S/N as high as 350 (200), but 
it can be clearly resolved by the next-generation spectrograph from 
a spectrum with S/N $\sim30$.

\begin{figure}
  \centering{
    \plotone{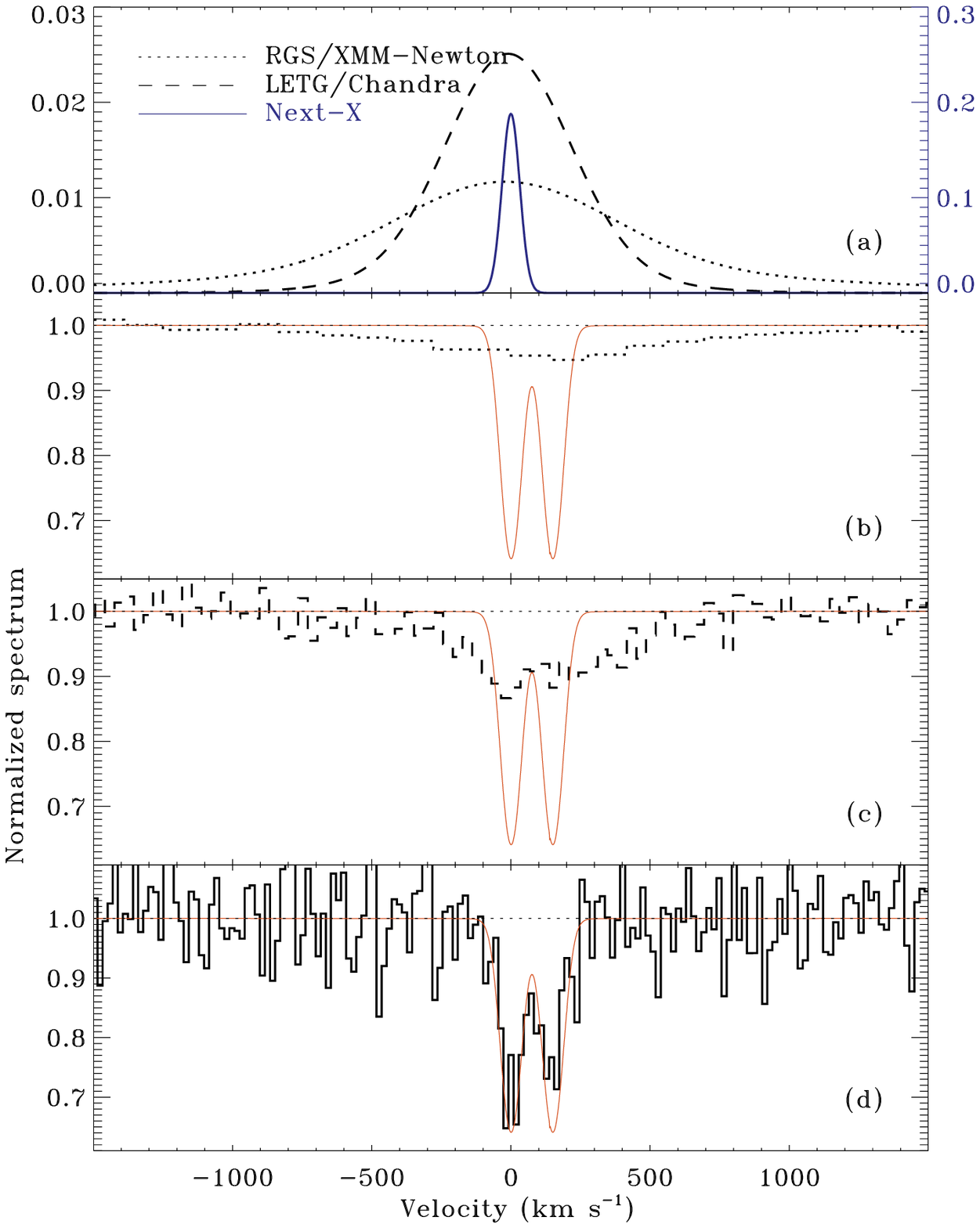}
  }
  \caption{Panel (a) shows line spread functions of \xmm-RGS (taken from 
  observation 0099280201), \chandra-LETG (taken from observation 4148),
  and a future spectrograph (denoted as ``Next-X'') approximated with a
  Gaussian with FWHM $=75~{\rm km~s^{-1}}$. The Next-X uses the
  right-hand scale. Panels (b)-(d) show the same absorption model (red
  curves) with two velocity components separated by $150~{\rm km~s^{-1}}$
  and both characterized as 
  $N_{\rm OVII}=10^{15}~{\rm cm^{-2}}$ and $b_v=50~{\rm km~s^{-1}}$ with
  corresponding simulated absorption spectra (histograms) of \xmm-RGS
  (b), \chandra-LETG (c), and Next-X (d) with S/N (per 50 m\AA\ for \xmm\
  and \chandra\ and per 5.4 m\AA\ for Next-X) $=350$,  200, and 30,
  respectively. }
  \label{fig:lsf}
\end{figure}

With $R=4000$, an effective area of $A\ge100~{\rm cm^2}$ is needed for
the spectrograph to systematically survey the WHIM \ovii\ along
15 sight lines within 6 months.
Assuming that all \ovii\ absorbers have $b_v=75~{\rm km~s^{-1}}$, it would
take 11.2 Ms for a spectrograph
with $R=4000$ and $A=100~{\rm cm^2}$ 
to finish surveying those 15 QSOs with $\gsim0.2$
mCrab (Section~\ref{sec:exp}) and
measure the intervening WHIM absorbers with 
$N_{\rm OVII}\sim10^{15}~{\rm cm^{-2}}$ 
  at $\ge4\sigma$ significance. Assuming a 70\% on-target timing
  efficiency, this is equivalent to a half year. To 
constrain the absorbers with 
$N_{\rm OVII}\sim10^{14.5}~{\rm cm^{-2}}$ with the
same amount of exposure time, $A=1000~{\rm cm^2}$ is required
(Figure~\ref{fig:ixo}).

In this section, we focused on the instrumental requirement particular for
the study of the WHIM. In fact, X-ray absorption/emission line diagnostics  
can also provide essential information for the physics of the ISM,
stellar coronae, 
black hole accretion and outflow, 
quasar accretion and feedback, 
etc., and remarkable progress has already been made with \chandra\ and \xmm\
observations (e.g., \citealt{yao06, tes08, mil06, arav07}). 
With an X-ray spectrograph equipped with $R\sim4000$ and 
$A\sim100~{\rm cm^2}$, we should be able to explore all these fields in
great detail.


\section{Summary}
\label{sec:dis}


1. We analyzed all available (as of 2011 April) grating data
   and obtained two coadded spectra with S/N$\sim$90 per 50-m\AA\
   resolution element from \chandra\ observations, 
   and S/N$\sim$190 from \xmm\ observations. Neither \chandra\ 
   nor \xmm\ observations support the existence of
   the two WHIM systems previously reported by \citet{nic05}.

2. We ran bootstrap simulations for the detecting limits of the current
   X-ray telescopes. We find that the reported EWs of the \ovii\ 
   K$\alpha$ at $z=0.011$ and $z=0.027$ should have been measured at 
   $\ge3.7\sigma$ and $\ge2.3\sigma$, in contrast to the fact that
   we only obtained upper limits to the EWs. 

3. According to the numerical simulations, the WHIM absorbers with 
   $N_{\rm OVII} \gsim10^{15}~{\rm cm^{-2}}$ could sample $>30-50\%$ of the
   \ovii-bearing baryons. To systematically survey the 15 QSO
   sight lines along which the IGM \ovi\ absorbers have been detected,
   future X-ray telescopes should be able to facilitate the WHIM study via
   the X-ray absorption line spectroscopy from background QSOs with
   fluxes of $\sim0.2$ mCrab to find $\ge10$ WHIM systems. 
   It takes impractical long 
   (341 Ms for \chandra\ and 804
   Ms for \xmm) exposure time for current X-ray telescopes to complete the
   survey. 

4. It would require $\sim11$ Ms on-target exposures for future
   X-ray spectrographs equipped with spectral resolution 
   $R\gsim4000$ and effective area $A\gsim100~{\rm cm^2}$ to 
   finish the survey.

\acknowledgements

We are grateful to Britton Smith for making his numerical simulation results
available to us before publication, and to Charles Danforth for extensive
discussion on this work. We thank the anonymous referee for the constructive
comments that help to greatly clarify the manuscript. 
This work is supported by NASA through ADP grant NNX10AE86G, and 
grant NNX08AC14G to the University of Colorado for data analysis and
scientific discoveries related to the Cosmic Origins Spectrograph on the
Hubble Space Telescope.


\end{document}